\newcommand{\ket}[1]{\lvert #1\rangle}
\newcommand{\braket}[2]{\langle #1 \lvert #2\rangle}

\def\C{{\@QC C}}
\def\@QC#1{\mathpalette{\setbox0=\hbox\bgroup$\rm}%
  {\egroup C$\egroup\rm\rlap{\kern0.4\wd0\vrule
  width 0.05\wd0 height 0.97\ht0 depth -0.01\ht0}%
  #1\bgroup}}

\documentclass[aps,twocolumn,raggedbottom,prl,showpacs,nobalancelastpage,amssymb,groupedaddress]{revtex4}
\usepackage{graphicx}
\usepackage{amsmath}
\usepackage{amssymb}
\include{nem}
\begin{document}
\title{Different types of integrability and their relation to decoherence 
in central spin models}
\author{B.\, Erbe and J.\, Schliemann}
\affiliation{Institut f\"{u}r Theoretische Physik, Universit\"at
Regensburg, 93053 Regensburg, Germany}
\date{\today}

\begin{abstract}
We investigate the relation between integrability and decoherence in central 
spin models with more than one central spin. We show that there is a 
transition between integrability ensured by Bethe ansatz and integrability 
ensured by complete sets of commuting operators. This 
has a significant impact on the decoherence properties of the system, suggesting that it is not necessarily 
integrability or non-integrability which is related to decoherence, 
but rather its type or a change from integrability to non-integrability.
\end{abstract}
 \maketitle

The Liouville-Arnol'd theorem states that if a system with $n$ degrees of 
freedom has $n$ involutive integrals of motion, which are functionally 
independent, its Hamiltonian equations of motion are solvable via 
quadratures \cite{Arnold}. Such a system is called integrable. 
Despite for huge effort, so far it has not been achieved to adapt the 
concept of integrability to the quantum mechanical framework satisfactorily. 
At the present time there are two commonly accepted definitions: A 
quantum mechanical system is called integrable (i) if there is a Bethe 
ansatz \cite{Faddeev} or (ii) if the system has a complete set of 
commuting operators (CSCO) \cite{Frahm} sharing ``suitable'' 
properties (to be further explained below). Note that the notion of 
integrability in classical mechanics does not require the solvability of the 
quadratures. In this sense both of the aforementioned approaches are in 
direct analogy with classical mechanics. 

In investigations mainly focused on the first type of integrability, 
evidence has been found that it is related to transport properties 
\cite{St06}, to quantum phase transitions \cite{Emary}, and to 
decoherence \cite{Lages05, Dahmen}. 
Here systems of the form
\begin{small}
\begin{equation}
H=H_c + H_{c \leftrightarrow b} + \ldots \text{further terms}
\end{equation}
\end{small}
have been considered, where $H_c$ denotes a central system and 
$H_{c \leftrightarrow b}$ a coupling term between the central system and a bath. 
Mainly two roads have been followed. On the one hand, the influence of 
chaotic or regular baths on the decoherence of the central system 
has been investigated \cite{Lages05}. On the other hand, 
the decoherence properties of the central systems of models which are 
integrable or non-integrable have been studied  
\cite{Dahmen}. The usual procedure within such 
considerations is to evaluate numerically the level statistics of the 
respective system and to relate a possible change in the statistics to a 
change of other properties of the system happening at the same point.

Motivated by their important role in the context of solid state quantum 
information processing \cite{SKhaLoss03}, 
we investigate in the present letter integrability and its relation to 
decoherence in central spin models. Here we define a quantum system to be 
integrable if it is possible to compute \textit{all} eigenstates and 
eigenvalues of the respective Hamiltonian using operations with less 
complexity than the direct diagonalization of the Hamiltonian matrix \cite{GMann}. 
Here we refer to the computional complexity. The exact diagonalization of
a Hamiltonian matrix for example grows exponentially with the system size.
This very strict notion of integrability contains (i) and (ii) as 
possible sources of integrability.

First we study the integrable structure of central spin models. 
In particular we show that there is a transition between integrability 
ensured by Bethe ansatz and integrability ensured by CSCO. 
Differently from the previous investigations described above, 
we then open a new route by applying a strong magnetic field to the 
central spin system, and analyze its reaction with respect to decoherence. 
In the non-integrable case as well as in the case of integrability ensured 
by Bethe ansatz, the strong magnetic field leads, as generally expected, 
to highly coherent central spin dynamics, whereas in the remaining case decoherence
still takes place. In contrast to 
previous work we relate the latter observation \textit{explicitly} to the 
\textit{type} 
of integrability and interpret the result from two different points of view.

The Hamiltonian of a central spin model is given by
\begin{small}
\begin{eqnarray}
\label{Ham}
\nonumber H&=&\sum_{i=1}^{N_c}\vec{S}_{i} \cdot 
\sum_{j=1}^{N} A^{i}_{j} \vec{I}_{j} + \sum_{i < j} J_{ij} \vec{S}_{i} \cdot \vec{S}_{j}\\
\nonumber &=& \left( \sum_{i=1}^{N_c} \vec{S}_{i}\right)  
\sum_{k=1}^N \left(   \frac{1}{N_c} \sum_{j=1}^{N_c} A^{j}_{k}\right) \vec{I}_{k} \\
\nonumber &+&\sum_{i=1}^{N_c} \sum_{j=i+1}^{N_c} \left(\vec{S}_{i}-\vec{S}_{j} \right) 
\sum_{k=1}^N \frac{1}{N_c} \left(A^{i}_{k}-A^{j}_{k} \right) \vec{I}_{k}\\
&+&  \sum_{i < j} J_{ij} \vec{S}_{i} \cdot \vec{S}_{j}, 
\end{eqnarray}
\end{small}
where in the following we consider $J_{ij}=J$ and $N_c>1$. 
For later convenience we define $A=N_c^{-1}\sum_{k=1}^N \sum_{j=1}^{N_c} A^{j}_{k}$. 
In the second identity we rewrote the original Hamiltonian into terms of 
sums and differences between the different central spins. 
The first term is nothing else than a Gaudin model \cite{Gaudin} with a 
central spin replaced by a sum over a set of spins, whereas the second 
term acts as a perturbation, vanishing whenever $A^{i}_{k}=A^{j}_{k}=:A_k$. Hence it has to be expected that this case is integrable, 
whereas the model generally should be non-integrable. This prediction has been 
verified explicitly in \cite{John09} by a detailed investigation of the 
spectral statistics of the model. We will come back to the integrable case
of two central spins with  $A^{1}_{i}=A^{2}_{i}=:A_i$ below.

Let us first, however, investigate in more detail general features of the
above system, fulfilling $A^{i}_{k}=A^{j}_{k}=:A_k$. 
The central spins can couple to different values of the total 
central spin squared $\vec{S}^2=\left(\sum_{i=1}^{N_c} \vec{S}_i\right)^2$. 
Fixing the associated quantum number $S$ and defining 
$\ket{0}=\ket{S}\ket{I_1, \ldots, I_N}$, we arrive at a usual 
Gaudin model with eigenstates \cite{Garajeu}
\begin{small}
\begin{equation}
\ket{N_D}=\prod_{i=1}^{N_D} \left(\omega_i S^- + \sum_{j=1}^N \frac{A_j\omega_i}{A_j-\omega_i}I_j^-\right)\ket{0}
\end{equation}
\end{small}
and eigenvalues
\begin{small}
\begin{equation}
E\left( \lbrace \omega_1, \ldots, \omega_{N_D} \rbrace  \right)= -2S\sum_{i=1}^{N_D} \omega_i+S\sum_{j=1}^N I_jA_j.
\end{equation}
\end{small}
The parameters $\omega_i$ are determined by the Bethe ansatz equations:
\begin{small}
\begin{equation}
S+\sum_{j=1}^N \frac{A_jI_j}{A_j-\omega_i}- 2 \sum_{k=1, k\neq i}^{N_D} \frac{\omega_k}{\omega_k-\omega_i}=0
\end{equation}
\end{small}
Here $N_D$ is the number of spin flips compared to $\ket{0}$ \cite{Garajeu}. 
Note that these equations are valid for any spin length $S$ and hence any number
of central spins $N_c$. Considering the Bethe ansatz equations instead
of the direct diagonalization of the Hamiltonian matrix reduces a problem of
exponential complexity to one of polynomial complexity. Hence the Hamiltonian (\ref{Ham})
with $A^{i}_{k}=A^{j}_{k}=:A_k$ is integrable, provided the Bethe ansatz equations
yield the correct number of solutions $\left\{\omega_1, \ldots, \omega_m \right\}$.
This however strongly depends on the inhomogeneity of the couplings $A_{k}$. 
Indeed for $A_k=(A/N) \Leftrightarrow A^{i}_{j}=(A/N)$,
the Bethe ansatz equations can never yield all eigenstates and eigenvalues.
 This becomes clear already on the subspace with only one spin flip. 
Here the Bethe ansatz equation becomes
\begin{small}
\begin{equation}
S + \frac{A}{A-N\omega}\sum_{j=1}^N I_j =0,
\end{equation}
\end{small}
which obviously gives only a single solution.
\begin{figure}
\begin{center}
\resizebox{0.55\linewidth}{!}{
\includegraphics{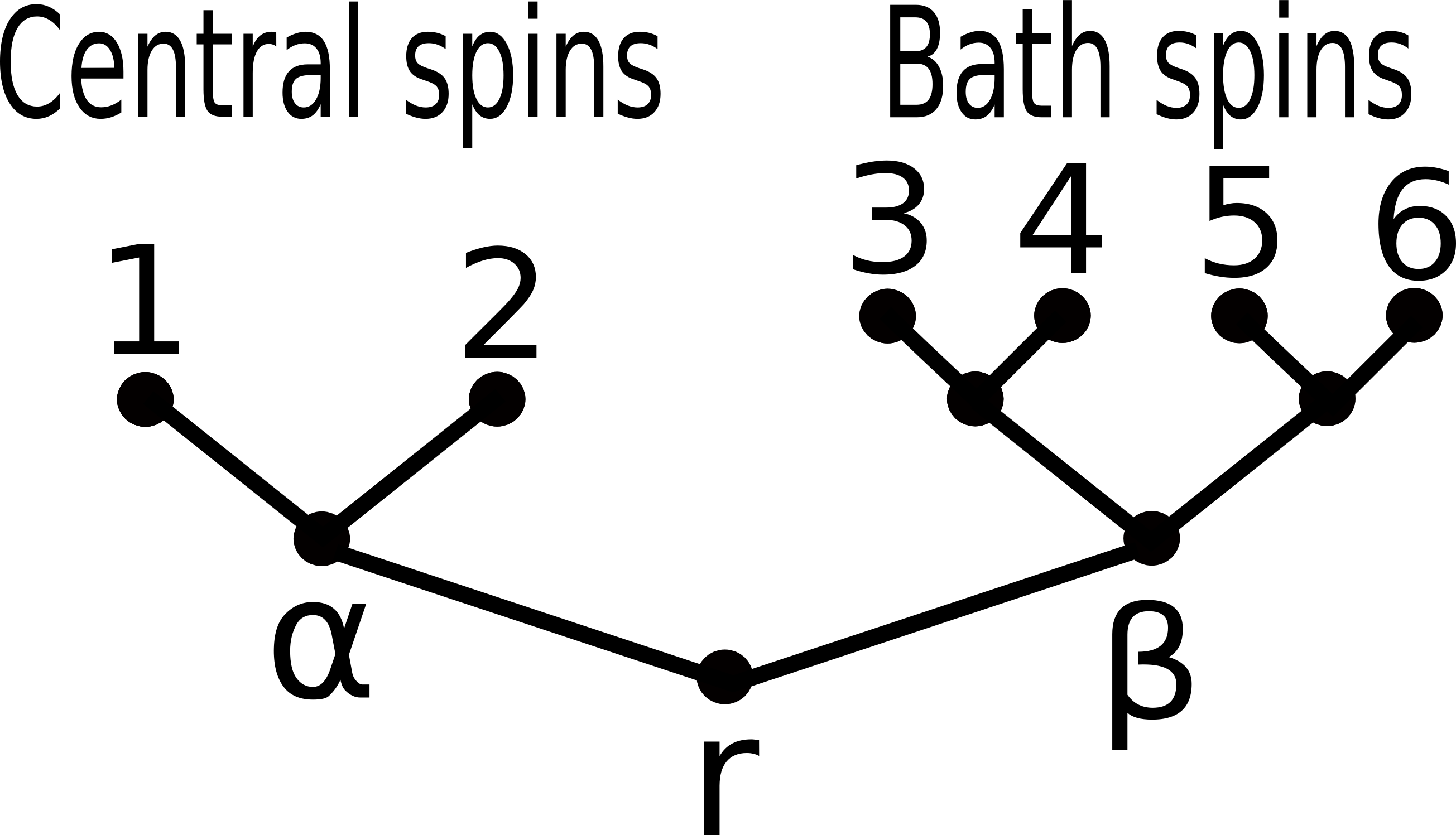}}
\end{center}
\caption{\label{Fig:Tree} Binary tree with $n=N_c+N=2+4=6$ leaves. 
In order to embed the Hamiltonian of an arbitrary central 
spin model with homogeneous couplings into a CSCO using the 
binary tree formalism, two binary trees $\alpha$, $\beta$ with 
$N_c$ and $N$ leaves respectively must be grafted together.
The Hamiltonian results as given in (\ref{inthom}).}
\end{figure}

Therefore integrability ensured by Bethe ansatz breaks if all couplings become 
identical. We now show that in this case integrability is ensured by CSCO. 
In order to construct the respective operators we apply the so-called binary tree 
formalism \cite{ErbS09}. On the first sight this seems to be unnecessary because Gaudin also 
gave the following set of operators which together with the Hamiltonian 
of his central spin model form a CSCO \cite{Gaudin}:
\begin{small}
\begin{equation}
 H_i = A_i \vec{S} \cdot \vec{I}_i - \sum_{j=1,j\neq i}^N \frac{A_i A_j \vec{I}_i \vec{I}_j}{A_i - A_j}
\end{equation}
\end{small}
Indeed these operators, which do not play any role concerning the construction 
of the eigenstates and eigenvalues of the Gaudin model, 
obviously become ill-defined in the homogeneous coupling limit. 
We restrict ourselves to a special case of
the binary tree formalism \cite{ErbS09} directly adapted to our model: Let $T$ be a binary tree with 
$n$ leaves as shown in Fig. \ref{Fig:Tree} for $n=6$. A binary tree consists 
of a set of nodes, each of which is connected to exactly two following nodes, 
except for the leaves. If we distinguish between a left and a right ``child'' 
$p_1$ and $p_2$ connected to a node $p$, we arrive at a natural ordering of 
the leaves. We denote the leaves a node $p \in T$ is connected to as $L(p)$. 
The node connected to all leaves is called the root, denoted by $r$ in the 
following. Now we associate every leaf $i$ with a spin $\vec{S}_i$ and define 
$H_p=\left( \sum_{i \in L(p)} \vec{S}_i\right)^2$ and $H_r^z=\sum_{i \in L(r)} S_i^z$. 
It is simple to see that for all $p \in T$ these operators commute. As every 
binary tree with $n$ leaves has $(n-1)$ nodes apart from the leaves, we thus 
arrive at exactly $n$ non-trivial, mutually commuting operators, which indeed 
form a CSCO. What makes these operators ``suitable'' in the sense of the 
introduction is the fact that they are complete for all spin lengths. In fact 
for any system it is possible to find a CSCO by e.g. considering the 
eigenbasis of the respective Hamiltonian and choosing a sufficient number of 
diagonal matrices with only one entry different from zero. We investigated such 
systems for the simple model of two Heisenberg coupled spins and found that 
they consist of more than two operators and lose the property of being complete, 
when the spin length is changed. We suppose that sets of commuting operators 
can only be complete for any spin length if the number of operators is equal 
to the number of spins. Surprisingly, up to our knowledge such a statement 
has not been made so far.

Now we show how to embed the Hamiltonian of an arbitrary central spin model 
with homogeneous couplings in a CSCO. To this end we consider two binary 
trees $\alpha$ and $\beta$ with $N_c$ and $N$ leaves respectively. Grafting 
them together as shown in Fig. \ref{Fig:Tree}, we arrive at a new binary 
tree with $N_c+N$ leaves. If we denote $\vec{I}_i$ as $\vec{S}_{N_c+i}$, the 
Hamiltonian of the associated homogeneous coupling model can be written 
in terms of elements of the CSCO resulting from the binary tree formalism as
\begin{small}
\begin{equation}
\label{inthom}
H=\frac{A}{N}(H_r-H_{\beta})+\left(\frac{J}{2}-\frac {A}{N} \right) H_{\alpha}.
\end{equation}
\end{small}
Note that the number of central and bath spins as well as their lengths are arbitrary and
that there is no further restriction to $\alpha$ and $\beta$ so that 
indeed there are numerous CSCO in which $H$ can be embedded. Furthermore, it 
should be mentioned that by adding $J_{b} H_{\beta}$ to (\ref{inthom}) we 
can easily include a homogeneous interaction of strength $J_b$ between the bath spins. It is simple to find the common eigenstates of the respective 
CSCO \cite{Steini09, ErbS09}:
\begin{small}
\begin{gather}
\nonumber  \ket{\lbrace S_{p \in T_L}\rbrace ,S^z_r} \\
\nonumber = \sum_{S^z_{p \in T_r} } \left(\prod_{p \in T_L} 
\braket{S_{p_1},S_{p_2},S_{p_1}^z,S^z_{p_2}}{S_p,S_p^z} \right) \\
\ket{S_1^z, \ldots, S_{N+2}^z},
\end{gather}
\end{small}
Here $T_L=T \setminus L(r)$, $T_r= T \setminus r$ and $S_p$ denotes 
the quantum number associated with $H_p$. The complexity for
calculating the Clebsch-Gordan coefficients is polynomial \cite{Loera} and hence
the approach indeed yields integrability. The eigenvalues read:
\begin{small}
\begin{eqnarray}
\nonumber E(\lbrace S_{p \in T_L}\rbrace ,S^z_r)&=&\frac{A}{N} 
\left(  S_r (S_r + 1)  - S_{\beta}(S_{\beta} + 1) \right) \\ 
&+& \left(\frac{J}{2}-\frac {A}{N} \right) S_{\alpha}(S_{\alpha} + 1)
\end{eqnarray}
\end{small}
Now we relate our above findings to the phenomenon of decoherence. 
The product of two spin operators consists of flip-flop terms involving ladder operators and
a coupling of the $z$-components \cite{BJ10}.
In the following we evaluate the dynamics for an initial state which is a
 simple product state. In this case all dynamics and hence all 
decoherence is purely due to the flip-flop terms. It is well-known that applying a magnetic field $B$ to the central spin 
system strongly suppresses the influence of flip-flop terms between the 
central spin system and the bath \cite{Coish10}. Here it is usually expected 
that whenever the magnetic field exceeds all other energy scales $B \gg |A|$,
 a complete neglect of their influence is justified. In the following we show 
that the effect of strong suppression of those flip-flop terms actually 
relies on the inhomogeneity of the couplings and is weakened stronger and stronger
the more couplings are chosen to be equal to each other.

\begin{figure}
\begin{center}
\resizebox{0.7\linewidth}{!}{
\includegraphics{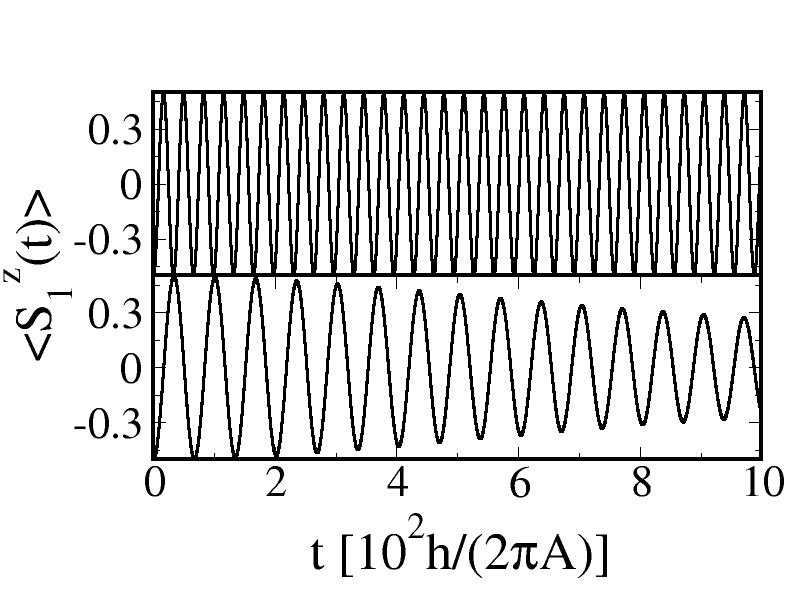}}
\end{center}
\caption{\label{Fig:hih} Spin dynamics for $N_c=2$, $N=2N_D+1=11$, where
$S_i=I_i=(1/2)$, and $B = 3.441A$,  $J = 0.023 A$.
The initial state of the system is 
$\ket{\Uparrow \Downarrow} \otimes \prod_{i=1}^{N_D} I^-_i \ket{0}$.
We consider two integrable models fulfilling $A^{1}_{i}=A^{2}_{i}=:A_i$
with either $A_i \neq A_j$ (upper panel), chosen due to a non-uniform distribution,
or $A_i=A_j$ (lower panel). Although in both cases $B$ is 
larger than any other energy scale, for homogeneous couplings the dynamics 
still decays, indicating decoherence.}
\end{figure}

To this end in Fig. \ref{Fig:hih} we consider the special case $N_c=2$ with $S_i=I_i=1/2$
and plot the spin dynamics for two integrable models ($A^1_j=A^2_j=:A_j$, as explained 
above) with inhomogeneous and 
homogeneous coupling costants. In the first case the coupling constants $A_j$ are chosen with
respect to a non-uniform distribution so that $A_i \neq A_j$. 
For our initial state this case can only be accessed via exact diagonalization, strongly restricting the 
size of the system \cite{SKhaLoss03}. We therefore illustrate the two situations considering a comparatively 
small system with $N=2N_D+1$ and $N_D=5$. This corresponds to a very low 
bath polarization of $1/N$. The initial state of the central spin system 
is $\ket{\Uparrow \Downarrow}$. We checked the dynamics for much larger 
systems in the homogeneous case using a semi-analytical approach based on \cite{BJ10} and did not find any qualitative differences. 
Moreover, non-integrable systems with fully inhomogeneous couplings
$A^i_k \neq A^j_k$ show a qualitatively very similar behavior to the
integrable case of inhomogeneous couplings, $A^{1}_{i}=A^{2}_{i}=:A_i$ and
$A_i\neq A_j$. Note that all results derived for the special case
of $N_c=2$ and $S_i=I_i=1/2$ in the following can be directly adapted to the general case
of an arbitrary number of central spins and arbitrary spin lengths. 

\begin{figure}
\begin{center}
\resizebox{0.6\linewidth}{!}{
\includegraphics{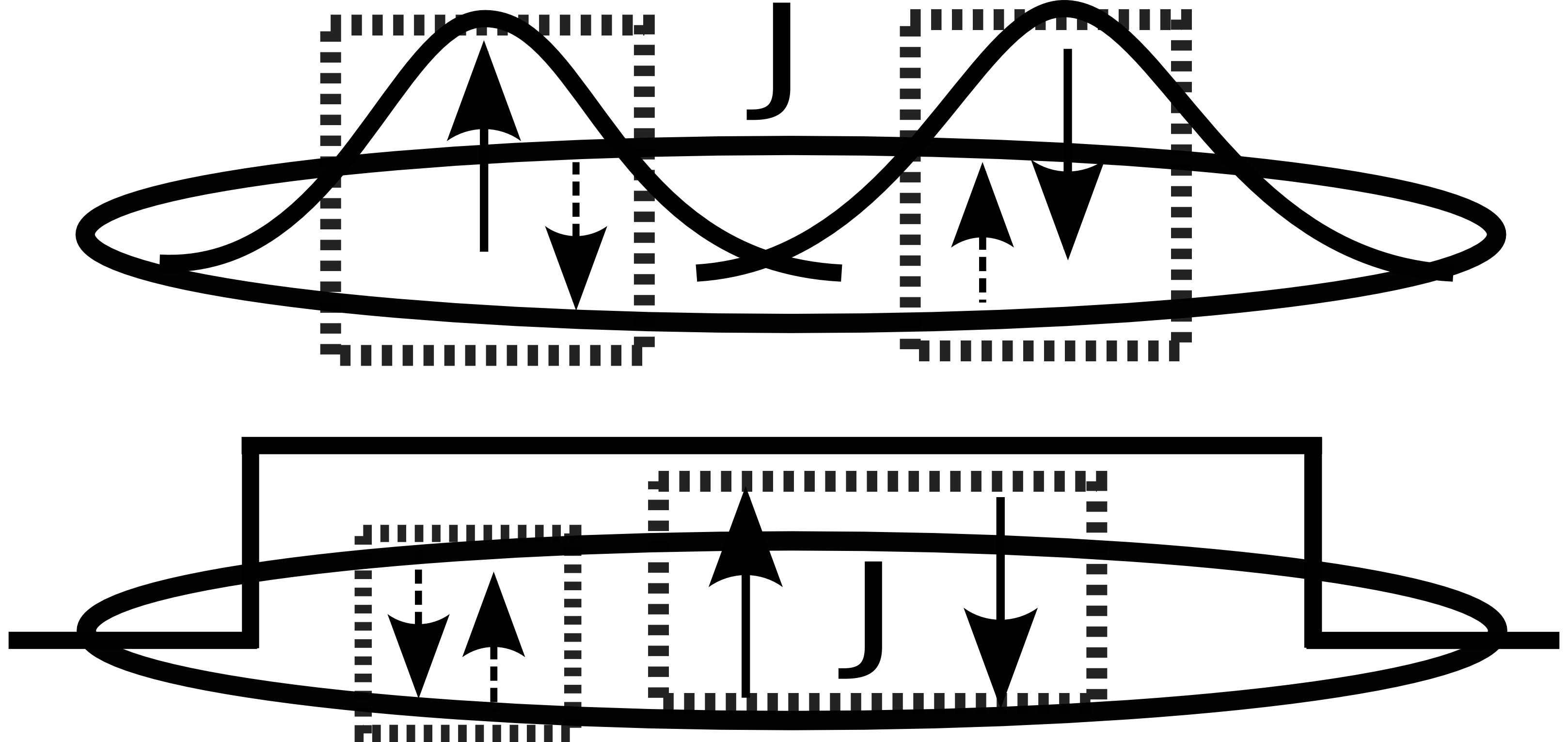}}
\end{center}
\caption{\label{Fig:FlipFlop} Sketch of flip-flop processes in a $N_c=2$ 
central spin 
system with inhomogeneous (upper panel) and homogeneous (lower panel)
couplings. For homogeneous 
couplings, the energy changes due to simultaneous central and nuclear spin 
flips can compensate each each other. This is not possible for inhomogeneous 
couplings because the energies of flips differ for different nuclear spins. }
\end{figure}

Although the magnetic field is in both cases 
larger than any other energy scale, the dynamics 
for the inhomogeneous case is completely coherent, whereas in the other case 
it still decays. This means that in the inhomogeneous case the flip-flop terms 
between the central spin system and the bath do not contribute to the dynamics 
in any determinable way. The oscillations are completely due to the flip-flop 
terms between the two central spins. 
A qualitative explanation of the above effect goes as follows:
Flipping a spin in a magnetic field changes the energy $E$ by 
$\Delta E \propto B$. In order to ensure energy conservation this change 
must be compensated. As indicated in the upper sketch of 
Fig. \ref{Fig:FlipFlop}, for inhomogeneous couplings this has to 
be done by the energy change due to the flip of the respective bath spin and 
the one resulting from the central spin flip via the central spin coupling term. 
Hence if the magnetic field exceeds any other energy scale, this is impossible 
and flip-flop processes are forbidden by energy conservation (at least in
first-order time-dependent perturbation theory). 
If we instead consider homogeneous couplings, this restriction can be 
circumvented by simultaneous flip-flop processes on both of the central spins. 
Here the energy changes due to the central spin flips in the 
magnetic field and the bath spin flips compensate each other as 
depicted in the second sketch of Fig. \ref{Fig:FlipFlop}. This is 
impossible for inhomogeneous couplings because the energy change depends on 
which bath spin is flipped. 

This simple effect vanishes for initial 
states with a fully polarized central or nuclear 
spin system. However, from the above explanation it is clear it will still 
occur if the couplings are varied away slightly from complete homogeneity. 
This means that the \textit{more} the couplings approach the CSCO integrable limit, the 
\textit{less} flip-flop terms are suppressed by a magnetic field applied to the central spin system. This leads to two 
different interpretations of the results, both of which indicate that it is 
not necessarily the integrability or non-integrability itself which is related to 
decoherence, as assumed in previous studies 
\cite{Lages05, Dahmen}: (a) As 
demonstrated above, the influence of a magnetic field applied to the central 
spin system on the decoherence properties strongly differs for models which
are clearly non-integrable or integrable by Bethe ansatz and those which are near to the CSCO integrable limit. 
In the first case the dynamics becomes highly coherent, whereas in the second case it still decays. 
This suggests that it is the mathematical 
structure ensuring integrability, which determines the reaction of a system
on an external quantity applied to the central system with respect 
to the decoherence properties rather than the integrability or non-
integrability itself. (b) An even more general interpretation results 
from the observation that if we apply a magnetic field to the central 
spin system, the non-integrable models as well as those integrable by 
Bethe ansatz keep the respective property, whereas it is lost in the CSCO case. 
Hence the result suggests that if a model is close to a limit
in which the integrability is broken by some external quantity 
applied to the central system, its decoherence properties will be
stronger affected than those of a system near to a limit with stable integrability. It is therefore 
the \textit{breaking} of integrability which has a negative 
effect on the decoherence properties and not the 
actual integrability or non-integrability.

Of course our results have to be regarded as a first indication into this direction and it 
would be desirable to check them for more general external quantities 
on a wider class of systems. As explained above, in (\ref{inthom}) we can easily add a term describing
an interaction between the different bath spins. Hence in an immediate next step 
it would be interesting to check for which types of bath terms the Bethe ansatz
integrability still holds and if we can find effects similar to those 
described in this paper. In this context see e.g. Ref. \cite{Lages05}.

\acknowledgments
We thank F. G\"{o}hmann for valuable discussions. 
This work was supported by DFG program SFB631.

\end{document}